\documentstyle[12pt]{article}

\newcommand{\news}{\setcounter{equation}{0}}
\newcommand{\be}{\begin{equation}}
\newcommand{\ee}{\end{equation}}
\newcommand{\bea}{\begin{eqnarray}}
\newcommand{\eea}{\end{eqnarray}}
\newcommand{\bean}{\begin{eqnarray*}}
\newcommand{\eean}{\end{eqnarray*}}
\font\upright=cmu10 scaled\magstep1
\font\sans=cmss12
\newcommand{\ssf}{\sans}
\newcommand{\stroke}{\vrule height8pt width0.4pt depth-0.1pt}
\newcommand{\Z}{\hbox{\upright\rlap{\ssf Z}\kern 2.7pt {\ssf Z}}}

\newcommand{\C}{{\rlap{\rlap{C}\kern 3.8pt\stroke}\phantom{C}}}
\newcommand{\R}{\hbox{\upright\rlap{I}\kern 1.7pt R}}
\newcommand{\CP}{\C{\upright\rlap{I}\kern 1.7pt P}}
\newcommand{\mt}{\widetilde m}
\newcommand{\cpone}{\CP$^1$}
\newcommand{\ie}{{\sl ie}\ }
\newcommand{\half}{\frac{1}{2}}
\newcommand{\quarter}{\frac{1}{4}}
\newcommand{\threeq}{\frac{3}{4}}
\newcommand{\efarg}{(\frac{2Kz}{L}|m)}
\newcommand{\cd}{{\cal D}}
\newcommand{\cs}{{\cal S}}
\begin{document}
\pagestyle{plain}
\title{
\begin{flushright}
{\normalsize DAMTP 94-77} \\
\end{flushright}
\vskip 20pt
{\bf Kink Chains from Instantons on a Torus} \vskip 20pt}
\author{Paul M. Sutcliffe \\[20pt]
{\sl Department of Applied Mathematics and Theoretical Physics} \\[5pt]
{\sl University of Cambridge} \\[5pt]
{\sl Cambridge CB3 9EW, England} \\[20pt]
{\sl email\  p.m.sutcliffe@amtp.cam.ac.uk}\\[10pt]}

\date{September 1994\\[20pt]}

\maketitle

\begin{abstract}
We describe how the procedure of calculating
approximate solitons from instanton holonomies may be extended to the
case of soliton crystals. It is shown how sine-Gordon kink
chains may be obtained from {\cpone}\ instantons on $T^2$.
These kink chains turn out to be remarkably accurate approximations to
the true solutions.
Some remarks on the relevance of this work to Skyrme
crystals are also made.

\end{abstract}
\newpage
\section{Introduction}
\news
The construction of Skyrme fields from instantons was introduced
a few years ago by Atiyah and Manton \cite{AM}. Not only has it has
proved useful in understanding the solitons of the Skyrme model, but
the general procedure of obtaining approximate solitons from higher
dimensional instantons has been shown to apply to several other systems.
However, for all the cases studied so far the mean soliton density
\hbox{$\bar\rho=(number\ of\ solitons)/(volume\ of\ space)$} is zero.
Of course, for soliton theories considered in non-compact space this
condition is a requirement for the configuration to have finite energy.
However, there are many situations of physical interest for which
$\bar\rho\neq 0$. Two examples of relevance to the work in this
paper are kinks of the sine-Gordon model on a circle, which describe
fluxons in long (but finite) Josephson junctions \cite{P}, and Skyrme
crystals, which are relevant for describing regions of high baryon
density such as inside a neutron star \cite{K}. By soliton crystal
we mean a configuration which is periodic in space and contains a
finite (non-zero) number of solitons in a unit cell. By definition
such a configuration (excluding the case of infinite period) will have
non-zero $\bar\rho$, and both the above mentioned examples may be
viewed as soliton crystals.\\

In this paper we shall concentrate on the case of the periodic sine-Gordon
equation, where the soliton crystal is a kink chain. It will be shown
in detail how an approximate kink chain may be constructed from an
instanton on the torus.
Since the exact kink chain solution is known we are also able to
make a detailed study of the accuracy of the instanton generated
approximation.
In the limit
in which the kink period tends to zero the instanton approximation
becomes exact. In the limit of an infinite kink period there are two
cases to consider, depending on whether the length of the torus in the
euclidean time direction is infinite or finite. In the first case we
reproduce the already known approximation obtained from instantons on \R$^2$,
and in the second case we find a new, and substantially more accurate,
approximation which decays exponentially, in agreement with the true
kink solution.\\

We view this work as a simple lower dimensional analogue of the
problem for Skyrme crystals, where an explicit exact solution is not known
and the instanton method may prove useful. Some remarks are made on
issues raised by the analysis here that are also relevant to the Skyrme
crystal.

\section{Kink Chains}
\news
In this section we shall briefly review some results for the
 periodic sine-Gordon
equation in one space dimension. The real sine-Gordon field $\phi(x,t)$
is taken to be periodic (mod $2\pi$) in the space variable $x$ with
period $L$ \ie
\be\phi(x+L,t)=\phi(x,t)+2\pi n
\label{percon}\ee
for some $n\in\Z$. The integrability of the model allows the
construction of explicit
multi-soliton wave train solutions using, for example, the inverse scattering
transform \cite{FM}. Fortunately we shall not need these
complicated solutions, involving vector $\Theta$-functions, since we
wish to consider the case of a static kink crystal. This is a time
independent kink chain, satisfying (\ref{percon}) with $n=1$, for which
the solution may be given simply in terms of a Jacobi elliptic
function. Explicitly, we consider a segment of the kink chain,
$x\in[0,L]$, with the boundary conditions $\phi(x=0)=0$ and
$\phi(x=L)=2\pi$. Then this segment contains precisely one kink (or
soliton). Since we are considering only static configurations the
energy per soliton is given by
\be{\cal E}=\int_0^L{1\over 16}(\partial_x\phi)^2+{1\over 8}
(1-\cos\phi)\ dx.
\label{enk}\ee
By completing the square in the above energy density we obtain, in the
usual way, a Bogomolny bound ${\cal E}\ge 1$ with equality if and
only if
\be\partial_x\phi=2\sin{\phi\over 2}.
\label{bog}\ee\\

The static sine-Gordon equation determined by (\ref{enk}) is
\be\partial_x^2\phi=\sin\phi
\label{sgeom}\ee
which can be integrated once to give
\be\partial_x\phi=2\sqrt{\sin^2({\phi\over 2})+{1-\mt\over\mt}}
\label{fint}\ee
where $\mt\in(0,1]$ is a constant of integration which is related
to the period $L$ in a way given below. Comparing (\ref{bog}) with (\ref{fint})
we see that the Bogomolny bound is attained only if $\mt=1$, which
corresponds (see below) to the limit $L\rightarrow\infty$. For $\mt\ne 1$
we see that the
Bogomolny bound is exceeded. A simple change of variable allows the
integration of (\ref{fint}) to give the kink chain solution
\be\phi=2\,{\rm arcsin}({\rm cn}({x-{L\over 2}\over\sqrt{\mt}}|\mt)).
\label{kcs}\ee
Here the kink is situated in the centre of the interval and we have
used the standard notation (see for example \cite{HMF})
that ${\rm cn}(x|m)$ denotes the appropriate Jacobi elliptic function
with argument $x$ and parameter $m$. The period of the elliptic
function must be equal to $L$ which gives the relation between $\mt$
and $L$
\be L=2\sqrt{\mt}\widetilde K
\label{rellm}\ee
where $\widetilde K$ denotes the complete elliptic integral of the
first kind corresponding to the parameter $\mt$ {\sl ie}
\be \widetilde K=\int_0^{\half\pi}\ {d\theta\over\sqrt{1-\mt\sin^2\theta}}.
\ee\\
Substituting the solution (\ref{kcs}) into the energy formula (\ref{enk})
gives the energy to be
\be{\cal E}={1\over\sqrt{\mt}}(\widetilde E -{1\over
2}(1-\mt)\widetilde K)
\label{eks}\ee
where $\widetilde E$ denotes the complete elliptic integral of the
second kind with parameter $\mt$. The energy tends to
the Bogomolny bound ${\cal E}=1$ as $\mt\rightarrow 1$, and is
strictly monotonic increasing as $\mt$ decreases, indicating
that there are repulsive forces between kinks.

Let us mention the two extreme limits of $L$. The first is the small
period limit $L\rightarrow 0$, which is also the weak coupling limit
of the theory since the gradient energy will dominate the potential
term. Hence the asymptotic limit of the solution will be the linear function
\be\phi\sim{2\pi x\over L}.
\label{zpl}\ee
The second is the infinite period limit $L\rightarrow\infty$, when the
solution reduces to the standard expression for the sine-Gordon kink
on \R. In this situation it is convenient to position the kink at
the origin $x=0$, which is simply achieved by shifting the $x$
coordinate by $-{L\over 2}$. Taking the limit $L\rightarrow\infty$
the solution (\ref{kcs}) takes the familiar form
\be\phi=4\,{\rm arctan(e}^x).
\label{sgk}\ee

\section{ CP$^1$ Instantons on T$^2$}
\news
In this section we study the \cpone\ $\sigma$-model defined on the torus
with euclidean metric.
We use the gauge field formulation, defined in terms of a 2-component column
 vector $V$, which is a function of the spacetime coordinates,
 $x^\mu=(x^1,x^2)=(x,y)$,
with metric $\delta_{\mu\nu}={\rm diag}(1,1)$
on $T^2$. The vector is
constrained to satisfy the normalization condition
\be V^\dagger V= 1.\ee
The action has a U(1) gauge symmetry and is given by
\be S=\int_{T^2}{\rm Tr}\ (D_\mu V)^\dagger(D^\mu V)\ d^2x \label{actcp}\ee
where Tr denotes trace and $D_\mu$ are the covariant derivatives
\be D_\mu=\partial_\mu - A_\mu \ee
with the composite gauge fields being purely imaginary and defined by
\be A_\mu=V^\dagger\partial_\mu V.\ee

It is convenient to also consider a gauge fixed formulation of the
model, obtained by introducing the parametrization
\be V={1\over \sqrt{1+\vert W \vert ^2}}\left(
\matrix{1\cr W\cr}\right)
\ee
where $W\in\ $\C$\cup\{\infty\}$. In terms of this
formulation the action (\ref{actcp})
becomes
\be S=\int_{T^2}{\partial_\mu W\partial^\mu\bar W\over
(1+|W|^2)^2}\ d^2x
\label{mactcp}\ee
and the Euler-Lagrange equation derived from (\ref{mactcp}) is
\be\partial_\mu\partial^\mu W={2\bar W\partial_\mu W
\partial^\mu W\over (1+|W|^2)}
.\label{cpoeom}\ee
We are interested in finite action solutions of this equation; which
in the language of differential geometry are harmonic maps from
$T^2$ to $S^2$, since \cpone\ is isomorphic to the Riemann sphere.
Such maps have an associated degree as described below.
Let $\zeta,\bar\zeta$ be coordinates on \cpone, and using the Fubini-Study
metric  construct the standard volume 2-form
\be{ \omega}={d\zeta\wedge d\bar\zeta\over (1+\zeta\bar\zeta)^2}.
\label{vtf}\ee
Then for a given map, $W:\ T^2\rightarrow$\ \cpone, the degree $(degW)$ is
determined by pulling back the volume 2-form (\ref{vtf}) to $T^2$ and
normalizing by the total volume of \cpone. Explicitly,
\be\int_{T^2}W^\ast\omega=(degW)\int_{\mbox{\scriptsize{CP}}^1}\omega.
\label{deg}\ee
Introducing the complex coordinate $z=x+iy$ on the torus
and using (\ref{vtf}) we obtain
\be degW={1\over\pi}\int_{T^2}{(|\partial_zW|^2-|\partial_{\bar z}W|^2)
\over (1+|W|^2)^2}\ dzd\bar z.
\label{top}\ee
We shall denote the degree $degW$ by the integer $N$, which in physics
is known as the topological charge or instanton number.
Returning to the gauge formulation it is easy to see that this integer
is precisely the (suitably normalized) magnetic charge
\be N={1\over i2\pi}\int_{T^2}F_{xy}\ d^2x\label{mag}\ee
where $F_{\mu\nu}=\partial_\mu A_\nu -\partial_\nu A_\mu$ is the
abelian field strength.\\

We now recall some results from differential geometry concerning the
harmonic maps we are interested in. The first theorem states
that \cite{EW}\\

{\sl Any harmonic map from a compact Riemann surface of genus $g$
to the two-sphere is holomorphic, provided its degree is greater than or
equal to $g$.}\\

Hence all the solutions we are interested in with positive
instanton number $N$ are given by holomorphic maps. (Those with
$N<0$ are given by anti-holomorphic maps). It is easily seen that
taking $W$ to be a holomorphic function of $z$ solves the equation
(\ref{cpoeom}). These are the so called instanton solutions.
It is clear that no solution can exist with $N=1$ since this would
imply that  $T^2$ and $S^2$ were diffeomorphic. The question of which
solutions exist is answered by the following theorem \cite{EL}\\

{\sl Holomorphic representatives for all maps from a Riemann
surface of genus $g$ to the two-sphere exist for degree greater than
$g$.}\\

Applying the above theorems to our case of interest we conclude that
all finite action solutions are instantons, which exist if and only if
$N\neq\pm 1$.\\

Let us now make a brief comment on the relevance of zeros and poles
of $W$. In the case of \cpone\ instantons on \R$^2$, finite energy
requires that the field tends to the same value at spatial infinity
irrespective of the direction of approach {\sl ie} there exists a
constant vacuum value $W_{vac}$ such that $W(z)\rightarrow W_{vac}$
as $|z|\rightarrow\infty$. Then $\widetilde W_{vac}=-\bar W_{vac}^{-1}$,
which is the point on the target manifold 2-sphere antipodal to
 $W_{vac}$, is the
value associated with the $W$ field at the position of an instanton.
That is, an instanton has position $z=z_0$ if $W(z_0)=\widetilde W_{vac}$.
Usually the vacuum value is taken to be $W_{vac}=0$ (or $\infty$) then
$\widetilde W_{vac}=\infty$ (or $0$) so that poles (or zeros) of $W$
are important since they are associated with the location of
instantons. For the instantons considered in this paper there is no
concept of a vacuum value since $T^2$ is compact. However, we shall
see that the poles and zeros of $W$ will still be important in
interpreting the location of an instanton. \\

We have already given a definition (\ref{deg}) of the degree of a map
from $T^2$ to $S^2$, but
an alternative (and of course equivalent) definition is also useful.
The degree of the map is equal to the number of preimages (counted with
the sign of the Jacobian of the mapping) of a given generic
image point in $S^2$. As we have seen the simplest case corresponds
to $N=2$ ($N=0$ is trivial and no solutions exist for $N=1$). So in searching
for a 2-instanton solution we require a holomorphic elliptic function
with two simple poles in a unit cell. Thus Jacobi elliptic functions
provide examples of 2-instanton solutions. It is such a solution which
we shall use in this paper. \\

The notation and results on elliptic functions that we shall use may
be found in \cite{HMF}. Let $m\in[0,1]$ be the elliptic function parameter
with $m^\prime=1-m$ the complementary parameter. Denote by $K$ and
$K^\prime$ the complete elliptic integrals of the first kind
corresponding to $m$ and $m^\prime$ respectively.
As mentioned above the locations of the zeros and poles of an
instanton solution will be of interest. For the Jacobi elliptic
functions there is a simple diagrammatic representation of these
points, as we now describe.
Let $\Omega$ denote the
set of letters  $\{s,c,n,d\}$. Arrange the elements of $\Omega$ in a
lattice in the complex $z$-plane (as shown below). Here the $s$ at the bottom
left of the lattice is placed at the position $z=x+iy=0$ and the lattice
spacing in the $x$ and $y$ directions is $K$ and $K^\prime$
respectively.
If $p$ and $q$ are any two distinct elements of $\Omega$,
then the Jacobi elliptic function $pq(z|m)$ has a simple zero at the $z$ value
occupied by a $p$ and a simple pole at the $z$ value occupied by a $q$.\\[15pt]
\vbox{
\hskip 130pt
$
\begin{array}{ccccc}
s & c & s & c & s\\
n & d & n & d & n\\
s & c & s & c & s\\
n & d & n & d & n\\
s & c & s & c &s
\end{array}$\\[10pt]

\noindent{{\sl Fig 1: Schematic representation of the zeros
 and poles of Jacobi elliptic
functions.}}\\[10pt]}

The symmetries that we shall later require of the instanton solution
(see section 4),
together with some simplifications made for ease of analysis, single
out the function $dn(z|m)$ as the instanton we require for our
purposes. This function has real period $2K$ and imaginary period
$4K^\prime$,
({\sl ie} $dn(z|m)=dn(z+2K|m)=dn(z+4iK^\prime|m)$ )
so that the grid covering the torus corresponds to
the first three columns in Fig 1. From our second definition of the
topological charge it is clear that $dn(z|m)$ describes a 2-instanton, since
the letter $d$ occurs twice in the first 3 columns of Fig 1.
In fact we wish to scale the torus so that it corresponds to the
rectangle $(x,y)\in[0,L]\times[0,\tau]$, with periodic boundary
conditions. Explicitly, we take the \cpone\ field to be given by
\be
W=a\ dn(\frac{2Kz}{L}|m)
\label{is}\ee
where $m$ is related to $\tau$ by $\tau=\frac{2K^\prime L}{K}$.
Here $a$ is a positive parameter, which we shall now show determines
how localized the instanton is around a zero of $W$; although it is
more complicated than a simple instanton scale or width.\\

 From Fig 1.
it is clear that $W$ has zeros at $z=\half L+i\quarter\tau,\
\half L+i\threeq\tau$ and poles at $z=i\quarter\tau,\ i\threeq\tau$.
Let us denote the topological charge density by $Q$; that is $Q$ is
the integrand in the expression (\ref{top}), so that the integral of
$Q$ over $T^2$ gives $N$. Then for the solution (\ref{is}) we find
\be
Q(x,y)=Q(z=x+iy)=\frac{4K^2a^2m^2}{\pi L^2}\frac{|sn\efarg cn\efarg|^2}
{(1+a^2|dn\efarg|^2)^2} .
\label{chden}\ee
Note that although $W$ is periodic on the torus $[0,L]\times[0,\tau]$, the
charge density $Q$ is in fact periodic on the half torus
$[0,L]\times[0,\half\tau]$. Hence, even though
 no $N=1$ instanton
solutions exist, we may at least think of each half of the torus as
\lq containing one instanton\rq . We can therefore restrict the
following analysis to one zero and one pole of $W$. \\

At the zero of $W$ given by $z=\half L+i\quarter\tau$, the topological
charge density (\ref{chden}) is
\be
Q_{\rm zero}=\frac{4m^\prime a^2K^2}{\pi L^2}
\ee
whereas at the pole $z=i\quarter\tau$ it is
\be
Q_{\rm pole}=\frac{4K^2}{\pi L^2a^2}.
\ee
Therefore we have that
\be
\frac{Q_{\rm zero}}{Q_{\rm pole}}=m^\prime a^4\label{zeropole} \ee
demonstrating that $a$ determines the extent to which the instanton
is localized around the zeros or poles of $W$. It is easy to confirm
this simple picture by plotting (\ref{chden}) for various values of
$a$. This shows that indeed for large $a$ the charge density is
concentrated in regions around the zeros of $W$ and for small $a$
it is concentrated in regions around the poles of $W$. For
intermediate values  $a\sim 1$ the charge density is not well
localized around merely zeros or poles but spreads out  along lines
joining zeros to poles.

\section{Instanton Holonomies}
\news
The idea of obtaining approximate solitons from instantons
was introduced by Atiyah and Manton \cite{AM} in the context of the
Skyrme model. They showed that calculating the holonomy
of an SU(2) Yang-Mills instanton in \R$^4$, along a line
parallel to the euclidean time axis, produces a Skyrme field
in \R$^3$ with baryon number (topological charge) equal to the
instanton number of the original Yang-Mills instanton. Moreover,
such Skyrme fields are very good approximations to the soliton solutions
(Skyrmions) of the theory. For example, with an appropriate choice
of instanton scale the energy of an instanton generated 1-soliton
is only 1\% higher than that of the (numerically) known solution.\\

Of relevance to the work in this paper is a lower dimensional analogue
of this result, that was introduced by the author in \cite{PMS}.
There it was shown that calculating the holonomy of a \cpone\ instanton
in \R$^2$, along a line parallel to the euclidean time axis, produces
a sine-Gordon kink field in \R\ with soliton number equal to the
instanton number of the original \cpone\ instanton. The fundamental
relation is
\be
\phi(x)=-i\int_{-\infty}^{+\infty}A_y(x,y)\ dy\ +N({\rm mod 2})\pi
\label{hol}\ee
which constructs an $N$-kink sine-Gordon field $\phi$
from the euclidean time
component $A_y$ of a \cpone\ $N$-instanton. Here we mention only the
$N=1$ result, since this will appear later as a limiting case of our
periodic construction. From a 1-instanton located
at the origin with scale $\lambda$ the formula (\ref{hol}) gives
\be
\phi=\pi(1+\frac{\lambda x}{\sqrt{1+\lambda^2x^2}}).
\label{ka}\ee
For some $\lambda$ this is meant to approximate the exact 1-kink
(\ref{sgk}). The energy of (\ref{ka}) is minimized at $\lambda=0.695$
where it takes the value ${\cal E}=1.010$ {\sl ie} only 1\% higher
than the exact solution. By closing the contour of integration in
(\ref{hol}) with a large semicircle at infinity and using Stokes\rq\
theorem one can see \cite{SW} that $\phi$ may also be expressed
as an area integral of the topological charge density. Explicitly,
\be
\phi(x)=2\pi\int_{-\infty}^{x} dx^\prime \int_{-\infty}^{+\infty}
dy\ Q(x^\prime,y)
\label{icd}\ee
where $Q(x,y)$ is the topological charge density of the \cpone\
instanton in \R$^2$.\\

We now describe how a periodic
analogue of this procedure exists. We construct approximations
to the kink chain solutions of section 2 using the instanton on $T^2$
described in section 3. We use the notation set up in these earlier
sections.\\

It will be convenient to begin from a charge density area integral,
of the form (\ref{icd}). On $T^2$ this becomes
\be
\phi(x)=\pi\int_0^x dx^\prime\int_0^\tau dy\ Q(x^\prime,y)
\label{icdb}\ee
where $Q$ is given by equation (\ref{chden}) for the instanton we
are considering. It is immediately clear that by construction
$\phi$ satisfies the periodic kink chain boundary conditions
$\phi(0)=0$ and $\phi(L)=2\pi$. Now let us convert this formula to
a holonomy-type expression, in order to tie up with the general theory
and for ease of calculation later. We need at least four coordinate
patches in order to cover $T^2$. However, the kink field has the symmetry
that if $\half L\leq x\leq L$, then $\phi(x)=2\pi-\phi(L-x)$. This is
one of the symmetries that determined the choice of the $W$ function
(\ref{is}); {\sl ie} we require  that the generated kink also has
 this symmetry. This
means that we need only calculate $\phi$ on the half interval
$0\leq x\leq\half L$. In the required region of the torus
$(x,y)\in[0,\half L]\times[0,\tau]$
we require only two patches. We take these to be
$U=\{(x,y)\in (\quarter L -\epsilon,\half L]\times [0,\tau]\}$ and
$\hat U=\{(x,y)\in [0,\quarter L +\epsilon)\times [0,\tau]\}$, where
$\epsilon$ is a small parameter with $0<\epsilon<\quarter L$. We can
in practice reduce the overlap between $U$ and $\hat U$ to a minimum
by considering their closure by setting $\epsilon=0$.
Note that $W$ (given by (\ref{is})) has no singularities in $U$ and
no zeros in $\hat U$.
In order to
consider holonomies we need to revert to the gauge theory formulation.
Recall that in this formulation $Q=\frac{F_{xy}}{i2\pi}$ is of course
gauge invariant. The fact that the integral of $Q$ over the whole
torus is non-zero represents an obstruction to finding a periodic
non-singular gauge potential $A_\mu$ over all $T^2$. We can however
find gauge potentials $A_\mu$ and $\hat A_\mu$ which are non-singular
when restricted to the patches $U$ and $\hat U$ respectively.
Explicitly we have
\be A_\mu=-i{\rm Re}(\frac{\bar W\partial_\mu W}{1+|W|^2})\ee
and
\be \hat A_\mu=-A_\mu|W|^{-2}\ee
where Re denotes the real part. On the overlap region between $U$ and
$\hat U$ these two gauge potentials are related by
\be
\hat A_\mu=A_\mu+i\partial_\mu\alpha\ee
where
\be \alpha(x,y)={\rm Im}(\log W)\ee
and Im denotes the imaginary part. Substituting these expressions
into (\ref{icdb}) in the various patches allows the integration over
$x^\prime$ to be performed to give the holonomy result

\[ \phi(x) = \left\{ \begin{array}{ll}
-i\half\int_0^\tau (\hat A_y(x,y) -\hat A_y(0,y))\ dy
& \mbox{if }\ 0\leq x\leq\quarter L \\
 & \\
\half(-i\int_0^\tau (A_y(x,y) - A_y(0,y))\ dy\ +\alpha(\half L,\tau)
-\alpha(\half L,0)) & \mbox{if}\ \quarter L< x\leq\half L
\end{array} \right. \]
\be \ee
Substituting the explicit expression  (\ref{is}) for $W$ in the
above and making a scale change of variable for $y$ gives the
following final form for $\phi$. Introducing the notation
\begin{eqnarray*}
\cd & = & dn(\frac{2Kx}{L}|m)\\
\cs & = &sn(y|m^\prime) \\
\Delta & = & (1-\cd^2)(\cd^2-m^\prime)\\
\Xi & = & (1-m^\prime\cs^4)((1+a^2\cd^2)-\cs^2(\cd^2+m^\prime a^2))^{-1}
\end{eqnarray*}
we have the rather complicated expression

\[\phi=\left\{ \begin{array}{ll}
\cd\sqrt{\Delta}\int_0^{2K^\prime}\Xi(\cd^2-m^\prime\cs^2)^{-1}\ dy
 & {\rm if}\ 0\leq x\leq \quarter L\\
& \\
\pi-a^2\cd\sqrt{\Delta}\int_0^{2K^\prime}\Xi(1-\cs^2\cd^2)^{-1}\ dy
& {\rm if}\ \quarter L < x\leq \half L .
 \end{array} \right. \]
\be \label{holfin}\ee
Remarkably, the above two parameter ($m$ and $a$) kink field provides
an excellent approximation to the kink chain solution for all $L$.\\

The two extreme cases $L\rightarrow 0$ and $L\rightarrow\infty$ can
be treated analytically. First we consider the simplest case of
$L\rightarrow 0$. \\

In section 2 we discussed the small period limit of the exact solution
and stated that in this limit the solution becomes linear ({\ref{zpl}).
It is not at all obvious that a linear function limit exists for our
instanton generated kink. However, we shall now show that indeed such
a limit exists and emerges as we take the limit $m\rightarrow 0$ in
our construction.  To discuss this case it is more convenient to
work with the integrated charge density formulation (\ref{icdb}).
In order to produce a field $\phi$ which depends linearly on $x$ we see that
we require the instanton topological charge density to resemble some
kind of extended plane wave. With this in mind we see from
(\ref{zeropole}) that for $m^\prime\sim 1$ we need to set $a=1$
to obtain $Q_{zero}=Q_{pole}$. This is a requirement if we wish
to obtain a plane wave which passes through the zeros and poles of
$W$. We now make use of the following leading order behaviour of the elliptic
functions and integrals as $m\rightarrow 0$
\begin{eqnarray*}
sn(z|m)&\sim&\sin(z)\\
cn(z|m)&\sim&\cos(z)\\
dn(z|m)&\sim&1\\
K&\sim&\frac{\pi}{2}\\
K^\prime&\sim&\half\log(\frac{16}{m}).
\end{eqnarray*}
Using these results in (\ref{icdb}) we find the leading order relation
for $\phi$ is
\begin{eqnarray}
\phi&\sim&\frac{\pi^2m^2}{16L^2}\int_0^xdx^\prime\int_0^{\frac{L}{\pi}
\log(\frac{16}{m})}
(\cos^2(\frac{\pi x^\prime}{L})\sinh^2(\frac{\pi y}{L})
+\sin^2(\frac{\pi x^\prime}{L})\cosh^2(\frac{\pi y}{L}))\ dy \nonumber\\
&\sim&\frac{\pi^2m^2x}{64L^2}\int_0^{\frac{L}{\pi}
\log(\frac{16}{m})} \exp(\frac{2\pi y}{L})\ dy \nonumber\\
&\sim&\frac{2\pi x}{L}.
\end{eqnarray}
Hence we have shown that in the limit in which the period $L$ tends to
zero the instanton generated kink approximation becomes exact.\\

We now consider the limit of infinite period
$L\rightarrow\infty$. First of all it is more convenient to centre the
soliton at $x=0$, by shifting the $x$ varaiable $x\rightarrow
x-\frac{L}{2}$. It turns out that the best parameter choice (in the
sense of minimizing the energy of the approximate kink)
is to take $m\rightarrow 1$ and
$a\rightarrow\infty$ as $L\rightarrow\infty$.
Explicitly, these limits are approached with the asymptotic forms
\begin{eqnarray}
m&\sim&1-16\exp({-bL}) \nonumber\\
a&\sim&\frac{\sqrt{\beta}}{4}\exp({\half bL})
\label{aa}\end{eqnarray}
where $b$ and $\beta$ are arbitrary positive constants. Note that for
these asymptotic forms we have that
\be
\frac{Q_{zero}}{Q_{pole}}=m^\prime a^4\sim \frac{\beta^2}{16}\exp({bL})
\ee
so that the topological charge density is localized around the zeros
of $W$; as it should be to obtain a kink which decays at spatial
infinity in the infinite period case.
We now use the leading order behaviour of the elliptic
functions and integrals as $m\rightarrow 1$
\begin{eqnarray*}
dn(x|m)&\sim&\mbox{sech}(x)\\
sn(y|m^\prime)&\sim&\sin(y)\\
K&\sim&\half\log(\frac{16}{m^\prime})\\
K^\prime&\sim&\frac{\pi}{2}.
\end{eqnarray*}
After substituting these expressions into the holonomy formula
(\ref{holfin}) the integral simplifies sufficiently to be calculated
explicitly. The result is
\be
\phi(x)=\pi(1+\frac{\beta\sinh bx\cosh bx}
{\sqrt{(1+\beta\cosh^2bx)(1+\beta\sinh^2bx)}}).
\label{ipia}\ee
Calculating the period of the torus in the euclidean time direction
we find it is simply related to the parameter $b$ via
$\tau=2\pi b^{-1}$. So the torus has finite extent in the euclidean
time direction if and only if $b$ is non-zero.
One expects  that the approximate kink (\ref{ka}) obtained from
instantons in \R$^2$ should appear from the periodic construction
if the torus is taken to have infinite period in both directions.
By the above remark the limit $\tau\rightarrow\infty$ corresponds to
the limit $b\rightarrow 0$. From (\ref{ipia}) it can be seen that
to obtain a well defined kink field in this limit requires
$\beta\rightarrow\infty$. If we take both these limits such that
the combination $b\sqrt{\beta}\equiv\lambda$ is finite, then
(\ref{ipia}) becomes
\be
\phi=\pi(1+\frac{\lambda x}{\sqrt{1+\lambda^2x^2}})
\label{kab}\ee
which is precisely the kink field (\ref{ka}) obtained from instantons
in \R$^2$. As mentioned earlier the energy of this
approximation is only around 1\% greater than the exact solution, even
though it decays only power-like as compared to the exponential decay
of the exact solution (\ref{sgk}). Note that in some sense this
approximation obtained from instantons in \R$^2$ is the worst possible
limit of the $T^2$ instanton construction. This is because the decay
behaviour of (\ref{ipia}) is
\be
\phi(x\rightarrow -\infty)\ {\sim}\
\frac{4\pi}{\beta}\exp({2bx})
\label{db}\ee
so that the approximation has an exponential decay (as does
the exact solution) provided $b\neq 0$ and $\beta$ is finite.
 It is only in the
$\tau\rightarrow\infty$ case that the exponential decay is lost
and replaced by a power-like decay. An indication of the parameter
values for which the approximation (\ref{ipia}) has minimum energy
(we are using this criterion to define the \lq best fit\rq\ )
can be obtained by comparing the decay behaviour (\ref{db}) with
that of the exact solution;
\be
\phi(x\rightarrow -\infty)\ {\sim}\
4\exp(x).
\label{dbx}\ee
This suggests parameter values of $b=\half$ and $\beta=\pi$.
Numerically minimizing the energy of (\ref{ipia}) we find the
parameter values $b=0.589$ and $\beta=1.86$, which are reasonably
close to those required to optimize the decay behaviour. At these
parameter values  the energy of the approximate kink is
${\cal E}=1.000216$ which is only $\frac{1}{50}\%$ above the exact
value ${\cal E}=1$. This approximation is therefore far superior
to the approximation (\ref{kab}). It is also a slightly better
approximation than any of those obtained in reference \cite{SW}
from instantons in \R$^d$ for arbitrary positive integer $d$. Moreover all
the approximations derived there had a power-like decay.

Having discussed the case $L\rightarrow 0$, where we have shown
that the instanton generated approximation with parameters
$a=1$ and $m\rightarrow 0$ becomes exact, and the case
$L\rightarrow\infty$ where we found the approximate kink with
parameters $a\rightarrow\infty$ and $m\rightarrow 1$ has energy only
$\frac{1}{50}\%$ above the true value, it is tempting to conclude
that the case of general $L$
interpolates between these two results. This in fact proves to be
correct, as can be verified by numerically integrating the holonomy
formula (\ref{holfin}). The results are that as $L$ increases from
$0$ to $\infty$ the parameter values which minimize the energy of the
kink approximation increase monotonically; in the case of $a$ it
is from $1$ to $\infty$ and in the case of $m$ it is from $0$ to $1$.
The percentage excess energy of the approximate kink over the true solution
also increases monotonically from $0\%$ to $\frac{1}{50}\%$.
Note that the variation of $m$ over its whole range is a very clear
indication that the shape of the torus plays a crucial role in
the accuracy of the approximation. As $L$ increases from
$0$ to $\infty$ the shape of the torus, as measured by the ratio
of the length of its sides $\frac{L}{\tau}$,  also increases from
$0$ to $\infty$.
\section{Conclusion}
\news
We have shown how the construction of approximate solitons from
instanton holonomies may also be applied to the case of a soliton
crystal. A detailed analysis shows that the
excess energy
of the periodic sine-Gordon kink obtained from an instanton on the torus is
around  $\frac{1}{50}\%$,  or
less, for all periods. This is much smaller than the corresponding
excess energy (of the order of $1\%$) of the
kink on the infinite line obtained from instantons
in the plane. A crucial ingredient in achieving this accuracy
is the ability to vary the length of the torus in the euclidean time
direction. \\

The results described here are encouraging for the
use of an instanton construction of the Skyrme crystal.
Particularly the fact that for the situation discussed
in this paper the accuracy of the approximation
is much better in the crystal case than in the previously considered
infinite line example. Some preliminary results have been obtained
for the Skyrme crystal and these will be described elsewhere
\cite{MS}.\\

\noindent{\bf Acknowledgements}\\
Many thanks to Nick Manton for useful discussions and the EPSRC
for a research fellowship.\\

\end{document}